\documentclass{aa}  
\usepackage{graphicx}
\usepackage{txfonts}
\usepackage{verbatim}
\begin{document}

\title{EX\,Lupi in quiescence}
   \subtitle{}

   \author{N. Sipos\inst{1} \and  P. \'Abrah\'am\inst{1}, J. Acosta-Pulido \inst{2}, A. Juh\'asz\inst{3}, \'A. K\'osp\'al\inst{4}, M. Kun\inst{1}, A. Mo\'or\inst{1} \and J. Setiawan\inst{3}
          }

   \offprints{N. Sipos}

   \institute{Konkoly Observatory of the Hungarian Academy of Sciences, P.O. Box 67, H-1525
	     Budapest, Hungary, 
              \email{niki@konkoly.hu}
\and
 Instituto de Astrof\'isica de Canarias, E-38200 La Laguna, Tenerife, Canary Islands, Spain
\and
Max-Planck-Institut f\"ur Astronomie, K\"onigstuhl 17, D-69117 Heidelberg, Germany	            
\and
Leiden Observatory, Leiden University, P.O. Box 9513, 2300RA Leiden, The Netherlands         } 
   \date{Received January, 2009; accepted June, 2009}

% \abstract{}{}{}{}{} 
% 5 {} token are mandatory
 
  \abstract
   {}
    {EX\,Lup is the prototype of EXors, a subclass of low-mass pre-main sequence stars whose  episodic eruptions are attributed to temporarily increased accretion. In quiescence the optical and near-infrared properties of EX\,Lup cannot be distinguished from those of normal T Tau stars. Here we investigate whether it is the circumstellar disk structure which makes  EX\,Lup an atypical Class II object. During outburst the disk might undergo structural changes. Our characterization of the quiescent disk is intended to serve as a reference to
study the physical changes related to one of EX\,Lupi's strongest known eruptions in 2008 Jan--Sep.} 
   {We searched the literature for photometric and spectroscopic observations including ground-based, IRAS, ISO and Spitzer data. After constructing the optical--infrared spectral energy distribution (SED), we compared it with the typical SEDs of other young stellar objects and modeled it using the Monte Carlo radiative transfer code \emph{RADMC}. A mineralogical decomposition of the 10\,$\mu$m silicate emission feature and also the description of the optical and near-infrared spectra were performed. }
   { The SED is in general similar to that of a typical T Tauri star, though above 7\,$\mu$m EX\,Lup emits higher flux. The quiescent phase data suggest low level variability in the optical--mid-infrared domain. Integrating the optical and infrared fluxes we derived a bolometric luminosity of 0.7\,L$_{\odot}$.  The 10\,$\mu$m silicate profile could be fitted by a mixture consisting of amorphous silicates, no crystalline silicates were found. A modestly flaring disk model with a total mass of 0.025\,M$_{\odot}$ and an outer radius of 150\,AU was able to reproduce the observed SED. The derived inner radius of 0.2\,AU is larger than the sublimation radius, and this inner gap sets EX\,Lup aside from typical T Tauri stars.}
   {}


   \maketitle
%
%________________________________________________________________

\section{Introduction}

In 2008 January the amateur astronomer A. Jones announced that EX\,Lupi, the
prototype of the EXor class of pre-main sequence eruptive variables, had brightened
dramatically (\cite{ref:jones_2008}). The quiescent phase brightness of EX\,Lup,
an M0\,V star in the Lupus\,3 star forming region, is V$\,{\approx}\,$13 mag.
During its unpredictable flare-ups its brightness may increase by
1-5 mag for a period of several months  (\cite{ref:herbig_1977}). In the last decades EX\,Lup produced a number of eruptions, the last one in 2002 (\cite{ref:herbig_1977}, \cite{ref:herbig_2001}, \cite{ref:herbig_2007}). During the present outburst, which lasted
until 2008 September (AAVSO International Database\footnote{www.aavso.org})
EX\,Lup reached a peak brightness of 8\,mag in January, 
brighter than ever before.

According to the current paradigm, eruptive phenomena in pre-main sequence stars
(FU\,Ori and EX\,Lup-type  outbursts) are caused by enhanced accretion onto the central
star. During these phases of the star formation process intense build-up of the stellar mass takes place.
(\cite{ref:hartmann_1996}). In this picture circumstellar matter must play a crucial role in the eruption mechanism. Surprisingly little is known about the
 circumstellar environment of the EXor prototype, EX\,Lup. While its infrared excess emission
detected by IRAS and ISO was attributed to a circumstellar disk (\cite{ref:grasvelazquez_2005}, 2005),
no detailed analysis or modeling of the disk structure has been performed so far. Such an analysis could contribute to the clarification of the eruption mechanism and also answer the long-standing open question of what distinguishes EXors from normal T\,Tau stars.
\cite{ref:herbig_2008} found no optical spectroscopic features which would uniquely define EXors, and
he also concluded that their location in the $(J-H)$ vs. $(H-K_s)$ color--color diagram
coincides with the domain occupied by T\,Tau stars.

In this work we search the literature and construct a complete
optical--infrared
spectral energy distribution (SED) of EX\,Lup representative of the quiescent phase. The SED will be modeled using a radiative transfer
code, and will be compared with typical SEDs of T\,Tau stars, in order to reveal
differences possibly defining the EXor class. During the recent extreme eruption EX Lupi
was observed with a wide range of instruments. Our results on the physical
parameters of the system can be used as a reference for evaluating the changes related to this
outburst.

%__________________________________________________________________

\section{Optical / infrared data}
\subsection{Optical and near-infrared photometry from the literature}

We searched the literature and collected available optical and near-infrared photometric observations of EX\,Lup obtained during its quiescent phase. We defined as quiescence periods when the source was fainter than 12.5 mag according to the visual estimates and  V-band magnitudes in the AAVSO International Database. A visual inspection of the DSS and the 2MASS images revealed no nebulosity surrounding the star, thus we concluded that measurements taken with different apertures can be safely compared. The query results are presented in Tab.\,\ref{tab:data1} and plotted in Fig. \ref{fig:sed}.

\begin{table}
\begin{minipage}[t]{\columnwidth}
\caption{Optical and infrared observations of EX\,Lup collected from the literature, obtained in its quiescent periods. References: (1) Bastian \& Mundt (1979); (2) Mundt \& Bastian (1980); (3) Herbig et al (1992); (4) The DENIS database (DENIS Consortium, 2005); (5) Glass \& Penston (1974); (6) Hughes et al. (1994); (7) 2MASS All-Sky Catalog of Point Sources (Cutri et al. 2003)}
\label{tab:data1}
\centering
\renewcommand{\footnoterule}{}  % to avoid a line before footnotes
\begin{tabular}{l l l c}
\hline \hline
 $\lambda$ [$\mu$m]    &Julian Date & Magnitude  & Reference\\
\hline
0.36 (U) &2443349.545 &$14.21$ &1  \\
         &2443349.572 &$14.2$  &1  \\
         &2443349.681 &$14.23$ &1  \\
         &2443349.717 &$14.25$ &1  \\
         &2443349.728 &$14.28$ &1  \\
         &2443350.653 &$14.22$ &1  \\
         &2443353.659 &$14.13$ &1  \\
         &2443895.85  &$14.52$ &2  \\
\hline
0.44 (B) &2443349.545 &$14.33$ &1  \\
         &2443349.572 &$14.36$ &1  \\
         &2443349.681 &$14.24$ &1  \\
         &2443349.717 &$14.36$ &1  \\
         &2443349.728 &$14.33$ &1  \\
         &2443350.653 &$14.20$ &1  \\
         &2443353.659 &$14.07$ &1 \\
         &2443894.85  &$14.52$ &2  \\
         &2443895.85  &$14.42$ &2  \\
         &2448896     &$14.21$ &3  \\
\hline
0.55 (V) &2443349.545 &$13.13$ &1  \\
         &2443349.572 &$13.19$ &1  \\
         &2443349.681 &$13.27$ &1  \\
         &2443349.717 &$13.27$ &1  \\
         &2443349.728 &$13.21$ &1  \\
         &2443350.653 &$13.14$ &1  \\
         &2443353.659 &$13.22$ &1  \\
         &2443894.85  &$13.20$ &2  \\
         &2443895.85  &$13.22$ &2  \\
         &2448896     &$13.03$ &3  \\
\hline
0.64 (R) &2448896     &$12.15$ &3  \\
\hline
0.79 (I) &2451295.81  &$11.109$&4 \\
\hline
1.25 (J) &2441474     &$9.76$  &5  \\
1.25 (J) &2445075     &$9.92$  &6  \\
1.235(J) &2451314.781 &$9.728$ &7  \\
1.221(J) &2451295.81  &$9.92$  &4  \\
\hline
1.65 (H) &2441474     & $9.04$ &5  \\
1.65 (H) &2445075     &$9.11$  &6  \\
1.662(H) &2451314.781 &$8.958$ &7  \\
\hline
2.2 (K)  &2441474     &$8.82$  &5 \\
2.2 (K)  & 2445075    &$8.78$  &6  \\
2.159 (K$_s$)& 2451314.781&$8.496$ &7  \\
2.144 (K$_s$)& 2451295.81 &$8.82$  &4 \\
\hline
3.4 (L) & 2441474     &$\ge8.7$&5  \\
3.4 (L) & 2445075     &$8.05$  &6  \\
\hline
4.8 (M) & 2445075     &$7.54$  &6  \\
\hline
\end{tabular}
\end{minipage}
\end{table}

\subsection{ESO 2.2/FEROS optical spectroscopy}
 Three spectra of EX\,Lup in quiescent phase were taken in 2007 July
with FEROS at the 2.2 m MPG/ESO telescope in ESO La Silla, Chile. FEROS has a
spectral resolution of 48000 and a wavelength coverage from 360 to 920\,nm
(\cite{ref:kaufer_1999}).
The data reduction was done using the online FEROS Data Reduction System implemented
on the telescope, that produced 39 one dimensional spectra of echelle
orders and a merged spectrum of the entire wavelength range.

With a cross-correlation method we computed the projected rotational
velocity and radial velocity of EX\,Lup in quiescence from the FEROS
spectra.
We cross-correlated the spectra with a numerical template
mask (\cite{ref:baranne_1996}). The mean
radial velocity is $-0.948\pm0.1$\,km\,s$^{-1}$. From the B$-$V color of EX\,Lup and by using a projected rotational velocity calibration for FEROS
(\cite{ref:setiawan_2004} we derived a $v \sin i = 4.4\pm 2$\,km\,s$^{-1}$.  Sections of the spectrum showing the most prominent emission lines are plotted in Fig. 2.

\subsection{NTT/SOFI near-infrared spectroscopy}

We have found near-IR spectra in the ESO Data Archive. The spectra were obtained on 2001 May 4 using SOFI at the ESO 3.5 m NTT telescope. The data were taken under the program 67.C-0221(A) (PI:D. Folha).
Two spectra were taken using the blue and red grisms (R~$\sim 1000$), which 
cover the ranges $0.95-1.64$ and $1.53-2.52$\,$\mu$m, respectively. In both cases 
the total
exposure time was 500\,s, divided into 4 exposures obtained following an
 ABBA nodding pattern. The data were reduced using our dedicated routines
developed under the IRAF environment. The data reduction included 
sky subtraction, wavelength calibration and frame combination.
Spectra of the G3V star HIP~78466 were used to correct for telluric absorption. 
A modified version of the Xtellcor software (\cite{ref:vacca_2003}) was used in this step. The spectrum is presented in Fig. 3.

\subsection{Infrared Astronomical Satellite (IRAS)}
EX\,Lup is included in the IRAS Point Source Catalog with detections at 12, 25 and 60\,$\mu$m, and with an upper limit at 100\,$\mu$m. The measurements took place during a quiescent phase of EX\,Lup. In order to refine these data and also try to obtain a 100\,$\mu$m flux density value, we re-analyzed the IRAS raw scans using the
IRAS Scan Processing and Integration tool (SCANPI) available at
IPAC\footnote{http://scanpiops.ipac.caltech.edu:9000/applications/Scanpi/}.
Depending on the wavelength, 6-8 scans crossed the object. The scans were
re-sampled, averaged, and a peak was located within 30$''$
of the position of EX\,Lup. Then a point-source template was fitted and
fluxes were calculated. The flux errors were dominated by the uncertainty
in estimating the background level. The fluxes were color-corrected in
an iterative way by convolving the source SED and the IRAS filter
profiles. The color-corrected fluxes and flux uncertainties can be seen in
Tab.~\ref{tab:data2}.

\begin{figure*}[t]
   \centering
   \includegraphics[angle=0, width=18cm]{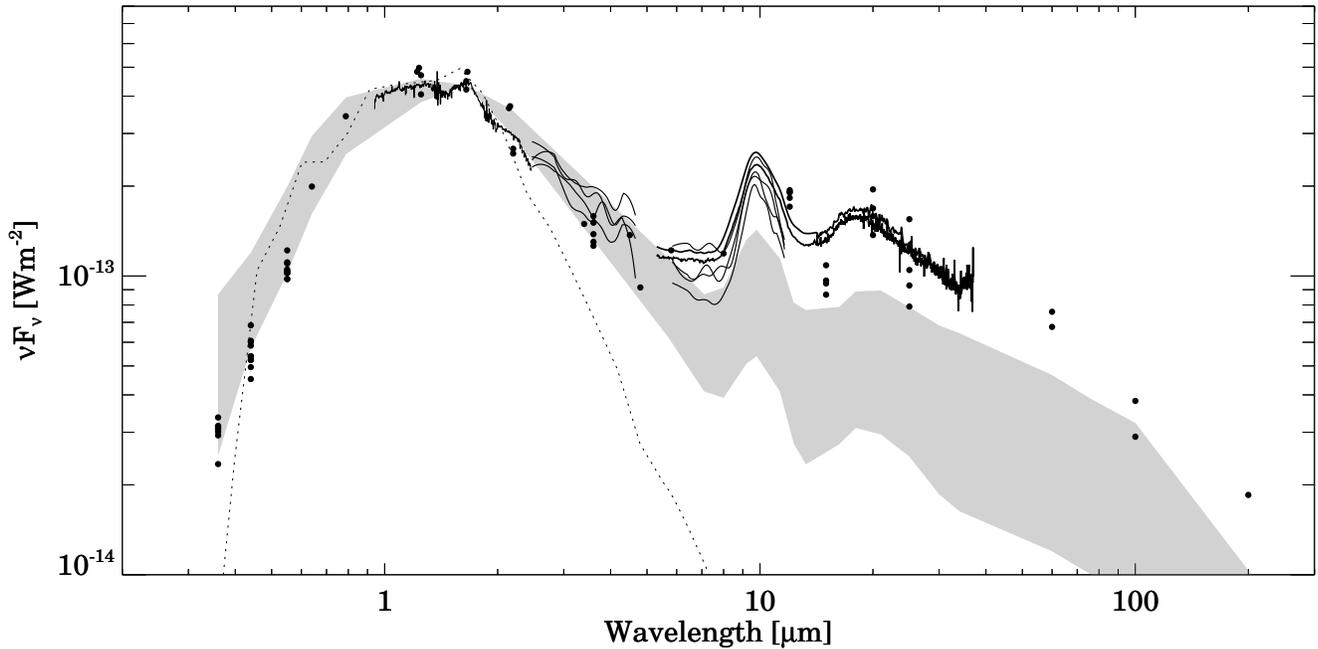}
      \caption{Spectral energy distribution of EX\,Lup, showing all data points from Tab. 1. and Tab. 2. and the smoothed spectra obtained with NTT/SOFI, ISOPHOT-S and Spitzer/IRS.  The gray stripe  marks the median SED of T Tauri objects from the Taurus-Auriga star-forming region.
The $5-36$ $\mu$m section of the median SED was constructed by \cite{ref:furlan_2006} based on Spitzer IRS data of 55 Class II objects with spectral types between K5 and M2, while the remaining part of our median SED was taken from \cite{ref:dalessio_1999} (1999) who computed their SED from 39 T Tauri stars. With dotted line we overplotted the stellar photosphere used in the models detailed in Sec. 4.3. }
         \label{fig:sed}
   \end{figure*}

\begin{table*}[ht!]
\caption{Color-corrected fluxes of EX\,Lupi. For the ISOPHOT data we adopted a conservative formal uncertainty of 15\%. }
\centering
\begin{tabular}{l c c c c c c}
\hline\hline
$\lambda$[$\mu$m] & 1983          & 1997 Feb 5  & 1997 Mar 18 & 1997 Aug 24 & 1997 Sept 19  & 2005 Mar 29 \\
&                   IRAS          & ISOPHOT      &ISOPHOT       &ISOPHOT       &ISOPHOT        &Spitzer/IRAC \\
\hline
3.6               &               &$0.17\pm0.03$  &$0.18\pm0.03$  & $0.15\pm0.02$  & $0.16\pm0.03$  &$0.190 \pm 0.004$\\
4.5               &               &               &               &                &                & $0.206\pm 0.004$ \\
5.8               &               &               &               &                &                & $0.236\pm 0.005$ \\
8.0               &               &               &               &                &                & $0.317\pm 0.007$\\
12                & $0.76\pm 0.1$ &$0.73\pm0.11$  &$0.78\pm0.12$  & $0.76\pm0.12$  & $0.68\pm0.11$  &\\
15                &               &$0.54\pm0.08$  &$0.43\pm0.07$  & $0.47\pm0.07$  & $0.48\pm0.07$  &\\
20                &               &$1.06\pm0.15$  &$1.30\pm0.18$  & $0.91\pm0.13$  & $1.12\pm0.16$  &\\
25                & $1.04\pm 0.1$ &$0.78\pm0.12$  &$0.87\pm0.14$  & $0.66\pm0.11$  & $1.29\pm0.21$  &\\
60                & $1.35\pm 0.4$ &$1.52\pm0.32$  &               &                &                &\\
100               & $0.97\pm 0.5$ &               &$1.27\pm0.17 $ &                &                &\\
200               &               &               &               & $1.23\pm 0.35$ &                &\\
\hline
\end{tabular}
\label{tab:data2}
\end{table*}

\subsection{Infrared Space Observatory (ISO)}

EX\,Lup was observed on 5 dates with ISOPHOT, the photometer
on-board ISO (\cite{ref:lemke_1996}). Four measurements belong to a small monitoring
programme which adopted similar instrumental setup at each epoch (PI: T. Prusti). For such datasets an algorithm aimed to achieve high relative photometry accuracy was developed in our group
(\cite{ref:juhasz_2007} 2007).
The raw data were corrected for instrumental effects following the scheme described in \cite{ref:juhasz_2007} (2007). Absolute flux calibration was performed by using the on-board
calibration source (FCS) for the 3.6--12\,$\mu$m filters, and by assuming a default responsivity for the 20--25\,$\mu$m range. At different single epochs 60, 100 and 200\,$\mu$m high resolution maps were
performed using the Astronomical Observing Template P32. These data were
 processed using a dedicated P32 Tool (\cite{ref:tuffs_2003}); the calibration was 
performed using the related FCS measurements.
The flux values were color-corrected by convolving the SED of EX\,Lup with
the ISOPHOT filter profiles. The results are listed in Table 2, the quoted formal uncertainties correspond to a conservative estimate of 15\%, however, the measurement error above 25~$\mu$m is probably higher than this value.
The fifth photometric dataset was published by \cite{ref:grasvelazquez_2005} (2005), but these data will not be used in our further analysis due to their presumably higher uncertainties.

At the four epochs of the monitoring programme also low-resolution spectrophotometry was performed
with the ISOPHOT-S sub-instrument in the 2.5$-$11.6\,$\mu$m wavelength range. These observations were reduced following the
 processing scheme described by
\cite{ref:abraham_2009}, which corrected for the slight
off-center positioning of the source, subtracted the separate
background spectra and adopted realistic error bars. The processed and
calibrated spectra are shown in Fig.\,1.

\begin{figure*}[ht]
\centering{
\includegraphics[width=16cm]{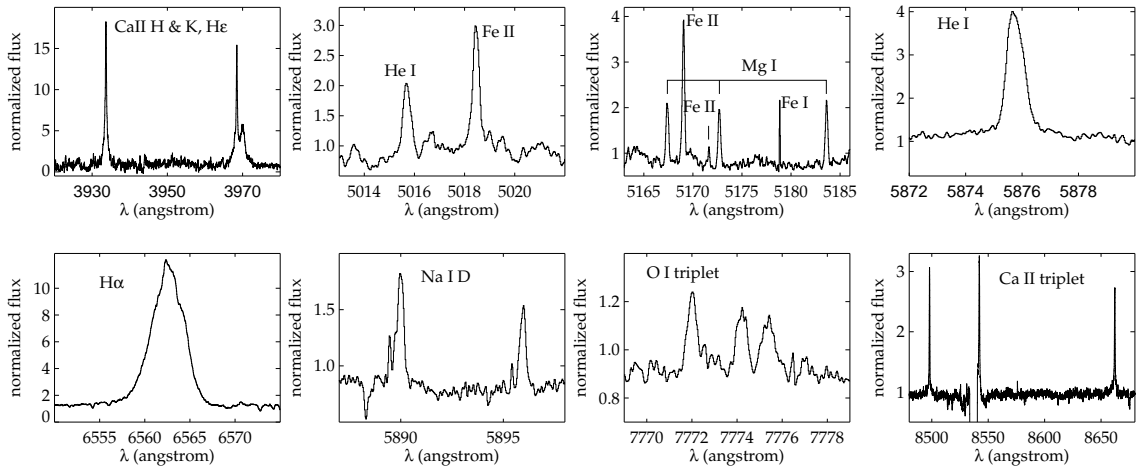}}
\caption{Details of the optical spectrum of EX Lupi showing significant or 
typical features. The interval shown in the second panel of the upper row is 
same as Herbig's (2007) Fig.~3. It is apparent that the 
broad component of the Fe~II~$\lambda$\,5018 line was absent in 2007. 
The third panel shows a typical region with narrow metallic lines. The 
O~I~$\lambda$\,7773 triplet, shown in the third panel of the lower row,
was observed in absorption in several T~Tauri stars (Hamann \& Persson 1992). }
\label{Fig_sp}
\end{figure*}

\subsection{Spitzer Space Telescope}

\paragraph{\textbf{\emph{IRAC.}}} EX Lup was observed with IRAC at 3.6, 4.5, 5.8 and 8.0\,$\mu$m
on 2005 March 29 (PID: 3716, PI: G. Stringfellow), in the sub-array mode. 
At each of the nine dither positions 64 images were obtained with  an exposure time of 0.1\,s.
The BCD (S14.0) image cubes of the 64 frames 
were combined into two-dimensional images using the "irac-subcube-collapse" 
IDL routine provided by the Spitzer Science Centre. 
Aperture photometry was performed on the nine final images
at each wavelength using a modified version of the IDLPHOT routines. The aperture
radius was set to 3 pixels (3.66\arcsec), the sky background was computed in an annulus
with an inner radius of 10 pixels and a width of 10 pixels (sigma clipping was
used to reject outlying pixel values).
The aperture correction was performed using the quoted values
from the IRAC Data Handbook (V3.0, hereafter IDH).
Following the outline of the IDH (see Chapter 5)  in the 3.6\,$\mu$m band a pixel-phase correction was also applied to the
measured flux densities.
The flux density values measured in the nine dither
positions (at each band) were averaged to get the final photometry.
To calculate the final uncertainty we added quadratically the
measurement errors (obtained from the nine individual flux density values in each band) and an absolute calibration error of 2\% (IDH).

\paragraph{\textbf{\emph{IRS.}}} 

EX Lup was observed with the Infrared Spectrograph (IRS, Houck et al. 2004) of the {\it Spitzer 
Space Telescope} on 2004 Aug 30 (PID: 172, PI: N. Evans) and on 2005 March 18 (PID:3716, 
PI: G. Stringfellow). On the first date the target was measured using Short Low (5.2--14.5\,$\mu$m), 
Long Low (14.0--38\,$\mu$m), Short High (9.9--19.5\,$\mu$m) and Long High (18.7-37.2\,$\mu$m)
modules. The integration time was 14\,s for the low-resolution modules, while 30\,s and 60\,s
were used for the Short High and Long High modules, respectively.
At the second epoch only the Short Low, Short High and Long High modules were used and
a PCRS peak-up was executed prior to the spectroscopic observation to acquire the target in the
spectrograph slit. The integration times was 14\,s for the Short Low module with 4 observing
cycles for redundancy, while 120\,s and 60\,s were used for the Short High and Long High modules, 
respectively, both with 2 observing cycles. 

The spectra are based on the {\tt droopres} and {\tt rsc} products processed through the S\,15.3.0 version
of the {\it Spitzer} data pipeline for the low- and high-resolution data, respectively. 
For the low-resolution spectra the background has been subtracted using
associated pairs of imaged spectra from the two nodded position, also eliminating stray light
contamination and anomalous dark currents. For the high-resolution spectra the background has
been removed by fitting a local continuum underneath the source profile. Pixels flagged by the
data pipeline as being "bad" were replaced with a value interpolated from an 8 pixel perimeter
surrounding the flagged pixel. The spectra were extracted using a 6.0 pixel and 5.0 fixed-width 
aperture in the spatial dimension for the Short Low and the Long Low modules, respectively, while
in case of the high-resolution modules the spectra were extracted by fitting the source profile
with the known PSF in the spectral images. The low-level fringing at wavelengths $>$20\,$\mu$m was 
removed using the {\tt irsfringe} package (\cite{ref:lahuis_2003}). The spectra are calibrated using a 
spectral response function derived from IRS spectra and MARCS stellar models for a suite of 
calibrators provided by the Spitzer Science Centre. To remove any effect of pointing offsets, 
we matched orders based on the point spread function of the IRS instrument, correcting for 
possible flux losses.
The spectra obtained at the two different dates agree within an uncertainty of $\le 11\%$, and are shown  in Fig. \ref{fig:sed}.

\section{Results}
\subsection{Variability in the quiescent phase}

The observations of EX\,Lup are sporadic, thus no complete SED could be constructed for any given epoch. We decided to merge all quiescent data regardless of their dates.
In Fig.~\ref{fig:sed} we plotted all
fluxes listed in Tabs.\,\ref{tab:data1} and \ref{tab:data2}. At most wavelengths the scatter of data points is in the order of 25\%,
though at optical and a few mid-infrared wavelengths the difference between the lowest and highest values can be considerably larger.    The ISOPHOT-S and Spitzer spectra shown in Fig.~\ref{fig:sed} also exhibit differences at similar level. Although part of this scatter is related to instrumental and calibration effects, the individual error bars are typically significantly smaller than the scatter of the data points, strongly suggesting the existence of an intrinsic variability. However, in case of the 60 and 100\,$\mu$m data points the uncertainty of the measurements is much higher than at shorter wavelengths, thus here the difference between the fluxes is within the error bars.  We consider the 25\% peak-to-peak variation as an
upper limit for the variability of EX\,Lup in quiescence. According to
Fig.~\ref{fig:sed}, the variability might be somewhat larger at
optical wavelengths, but the low number of data points prevents
us from claiming a wavelength-dependence.

\begin{table}
\begin{centering}
\caption{Observed wavelength, EW, and FWHM of the most prominent 
emission lines in the optical spectrum of EX Lupi in quiescence$^*$}
\label{Tab_sp}
\begin{tabular}{l c r r }
\hline
\hline
\noalign{\smallskip}
Line & $\lambda$ & EW & FWHM  \\
     &  (\AA) &  (\AA)~ &  (km\,s$^{-1}$) \\
\noalign{\smallskip}
\hline 	
\noalign{\smallskip}
Ca II H     &  3933.66   &   $-$13.09  &  45.8  \\
Ca II K     &  3968.46   &    $-$5.41  &  22.7   \\    
H$\delta$   &  4101.70   &   $-$14.29  &  155.1   \\	 
H$\gamma$   &  4340.42   &   $-$13.02  &  137.0   \\	 
H$\beta$    &  4861.25   &   $-$18.74  &  144.0   \\	  
He I (11)   &  5875.75   &    $-$1.96  &  37.6  \\     
Na I (1)    &  5889.96   &    $-$0.52  &  22.9   \\	 
Na I (1)    &  5895.94   &    $-$0.34  &  18.3   \\	 
H$\alpha$   &  6562.56   &   $-$35.95  &  191.0   \\	 
HeI (46)    &  6678.24   &    $-$0.67  &  23.8   \\	
He I (10)   &  7065.31   &    $-$0.42  &  29.3   \\    
Ca II (2)   &  8498.01   &    $-$1.51  &  23.0   \\	
Ca II (2)   &  8542.08   &    $-$2.26  &  40.5   \\    
Ca II (2)   &  8662.17   &    $-$1.49  &  27.0   \\	
\noalign{\smallskip}
\hline
\end{tabular}
%\end{centering}

\medskip
$^*$The full list of the emission lines identified in the spectrum
are found in the electronic version of this Table.

\end{centering}
\end{table}

\subsection{The optical--infrared SED}

The measured SED of EX\,Lup is presented in Fig.\,\ref{fig:sed}.
The optical part is clearly dominated by the stellar photosphere. The optical and near-infrared
colors, however slightly differ from the standard colors of an M0\,V star. 
This fact was already mentioned by \cite{ref:grasvelazquez_2005} (2005) who could not derive a positive extinction value from the $E_{B-V}$ and $E_{R-I}$ colors. Similarly \cite{ref:herbig_2001} claimed it to be unknown as well.
An infrared excess above the photosphere is detectable 
longwards of the K-band. The 3--8\,$\mu$m range is smooth and devoid of any
broad spectral features, indicating that EX Lup is neither deeply embedded to exhibit ices, nor hot enough for exciting PAHs. At 10\,$\mu$m a strong silicate emission appears, and a
 corresponding -- though  broader and shallower -- silicate band can be
seen around  20\,$\mu$m. At longer wavelengths the continuum
 emission decreases following a power law with a spectral index
of about $-4/3$. Due to the lack of any sub-mm
or mm measurement, the SED cannot be followed longwards of 200\,$\mu$m.

\subsection{Spectroscopy}

\subsubsection{The optical spectrum}

The high-resolution optical spectrum of EX Lupi, obtained on 2007 July 30,
shows a large number of emission lines in addition to the photospheric 
absorption features characteristic of the young late-type star.
The  emission lines are mostly symmetric, without noticeable 
P Cygni or inverse P Cygni absorption. The most prominent 
emission lines are the Balmer lines of the hydrogen, the H and K lines 
and the infrared triplet of the ionized calcium, and the helium 
lines at 5875 and 6678~\AA. The observed wavelengths and equivalent 
widths in \AA, and the full widths at  half maximum 
in km\,s$^{-1}$, determined by gaussian fitting,
are listed in Table~\ref{Tab_sp}. In addition to these prominent features, 
more than 200 weak, narrow metallic lines could be identified using  
Moore's (1945) multiplet tables. The widths of the lines are around 
10--20\,km\,s$^{-1}$, indicating that they originate from the active 
chromosphere of the star (Hamann \& Persson 1992).
All the emission lines identified in the spectrum are 
listed in the electronic version of  Tab.~\ref{Tab_sp}.  
Fig.~\ref{Fig_sp} shows sections of 
the spectrum.

The shape of the H$\alpha$ line is nearly symmetric. Its equivalent width,
W(H$\alpha$)=$-$35.9\,\AA, is significantly larger than the upper
limit of the chromospheric H$\alpha$ emission of M0 type stars
($\sim$6\,$~\AA$, Barrado y Navascu\'es \& Mart\'{\i}n 2003), indicating 
active accretion during the quiescent phase. The velocity width of the 
H$\alpha$ line 10\% above the continuum level is 362\,km\,s$^{-1}$, 
larger than the lower limit of  270\,km\,s$^{-1}$, set by White \& Basri 
(2003) for accreting T Tauri stars.
The empirical relationship between the 10\% width of the H$\alpha$ line
and the accretion rate $\dot{M}_\mathrm{acc}$, found by Natta et al. (2004), 
allowed us to derive 
$\dot{M}_\mathrm{acc} = 4.2_{-2.3}^{+8.1}\times10^{-10}$\,M$_{\sun}/$yr.
Comparison of the  H$\alpha$ line with that published by Reipurth,
Pedrosa \& Lago (1996) shows that in 2007 the H$\alpha$ emission line 
of EX Lupi was more symmetric, narrower, and had somewhat larger equivalent 
width than in 1994 March, a few days after an outburst.

\subsubsection{Near-infrared spectrum}

\begin{figure}
\centering{
\includegraphics[width=9cm]{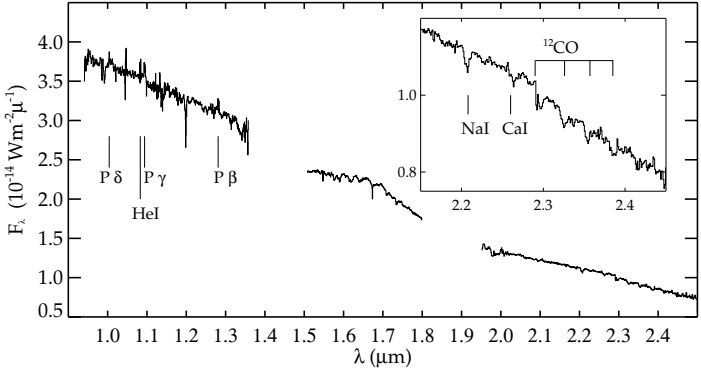}}
\caption{Near-infrared spectrum of EX Lupi, observed on 2001 May 4.  
The region around the $^{12}$CO bandhead absorption   
is enlarged in the inset.}
\label{Fig_irsp}
\end{figure}

The flux-calibrated infrared spectrum, observed on 2001 May 4, is displayed in 
Fig.~\ref{Fig_irsp}. The shape of the spectrum is similar to that of an 
unreddened M-type star without near-infrared excess (Greene \& Lada 1996). 
In the ZJ band the Paschen $\beta$, $\gamma$, $\delta$, and the 
He~I~$\lambda$\,1.083\,$\mu$m lines are seen in emission.
Absorption features of the Na~I at 2.21~$\mu$m,
Ca~I at 2.26~$\mu$m, and the rovibrational $\Delta \nu = +2$ transitions 
of CO at 2.3--2.4\,$\mu$m can be identified in the
K-band part of the spectrum. The Br$\gamma$ line of the hydrogen is
barely visible in emission, but its equivalent width cannot be
measured. The following absorption features are also identified in our spectrum:
at 1.199\,$\mu$m a very deep feature corresponding to S~I; at 1.577\,$\mu$m a
feature  associated to Mg~I and at 1.673\,$\mu$m another feature associated with
 Al~I.

The Pa$\beta$ emission line is a useful accretion tracer: its luminosity 
 correlates well with the accretion luminosity (e.g. Muzerolle et al. 1998,
Dahm 2008). We used the relationship established by Muzerolle et al. to 
derive the accretion luminosity of EX Lup during quiescence from the 
measured flux F(Pa$\beta$)=$1.7\times10^{-17}$\,W\,m$^{-2}$. The resulting 
$L_\mathrm{acc}=3.7\times10^{-3} L_{\sun}$ allows us to determine the
mass accretion rate $\dot M_\mathrm{acc}$, taking the mass and the radius 
of the star from Tab. 6. The accretion was assumed to proceed through a gaseous disk; for its inner radius we adopted  $R_{in} \approx 5\,R_{*}$ (Gullbring et al. 1998).
The result is 
$\dot M_\mathrm{acc} \approx 4.0 \times 10^{-10}$\,M$_{\sun}$/yr,
in good agreement with $\dot M_\mathrm{acc}$ obtained from the 
velocity width of the H$\alpha$ line. 
 A flux calibration uncertainty of 20\%  and the uncertainty 
of the empirical relationship sets the lower and upper limit at 
$1.3 \times 10^{-10}$\,M$_{\sun}$/yr and $1.1 \times 10^{-9}$\,M$_{\sun}$/yr,
respectively.

\section{Discussion}

\subsection{Comparison with young stellar objects}

Infrared SEDs of eruptive young stars (EXors and/or FUors) were examined by \cite{ref:green_2006}, \cite{ref:quanz_2007} (2007) and \cite{ref:agi_2009}. Some  of these stars (eg. UZ Tau, VY Tau, DR Tau, FU Ori, Bran 76) resemble EX\,Lupi: their SEDs decrease towards longer wavelengths and show a silicate feature in emission. The SEDs of the other group of FU Ori-type or EXor-like variables (eg. V1057 Cyg,  V1647 Ori, PV Cep, OO Ser), exhibit flat or increasing SEDs in the 20$-$100\,$\mu$m wavelength range (\cite{ref:abraham_2004}, \cite{ref:agi_2007}), which is clearly different from the shape of EX Lupi's spectral energy distribution. \cite{ref:green_2006} and \cite{ref:quanz_2007} (2007) suggested that the diversity in the shape of the SEDs is related to the evolution of the system: 
younger objects, still embedded in a large envelope exhibit flat SEDs, while more evolved objects, emitting T Tauri-like SEDs  have already lost their envelopes and have only circumstellar disks. According to this categorization we conclude that EX\,Lupi is relatively evolved among eruptive stars, and expect that its  circumstellar environment consists only of a disk without an envelope.

In order to compare EX\,Lup with other -- not eruptive -- young stars, in Fig. \ref{fig:sed} we overplotted a gray stripe  marking the median SED of T Tauri objects from the Taurus-Auriga star-forming region (\cite{ref:dalessio_1999} 1999, Furlan et al. 2006).
At shorter wavelengths the SED of EX Lupi follows the Taurus median. Longwards of approximately 7\,$\mu$m, however, its absolute level becomes higher than the median by a factor of $\sim$2.5. Nevertheless, even in this longer wavelength range its shape still resembles the Taurus slope. 

A further comparison of EX\,Lupi with other pre-main sequence stars can be based on the mid-infrared spectrum. \cite{ref:furlan_2006} presented a large sample of Spitzer IRS spectra of young stars and grouped them into several categories by measuring the slope of the SEDs and the strength of the 10 and 20\,$\mu$m silicate features. A visual classification places EX\,Lup in their scheme in group A or in group B, defined by a relatively strong silicate feature and a flat or somewhat decreasing SED over 20\,$\mu$m. \cite{ref:furlan_2006} suggested that only limited dust-growth and settling has taken place in these groups. Nonetheless, we found that EX\,Lup is not a typical  member of these groups, since shortwards of 8\,$\mu$m  the steep decrease of excess emmission characteristic of their objects is missing in the case of EX\,Lupi. These qualitative conclusions can be verified by calculating color indices for EX\,Lup. The strength of its silicate feature $(F_{10}-F_{cont})/F_{cont}=0.58$ is an intermediate value between group A and group B objects. However, the $n_{6-25}$ spectral index is typically negative for all objects in the sequence while positive (0.03) in the case of EX\,Lupi, making it an 'outlier' in the scheme. Also, the correlation between the $n_{6-13}$ and $n_{13-25}$ indices found by \cite{ref:furlan_2006} does not seem to hold for EX\,Lup. The colors of IRAS 04385+2550, a star exhibiting IRS spectrum very similar to that of EX\,Lup, were interpreted by \cite{ref:furlan_2006} as indication for the opening of an inner gap in the disk. This might also give an explanation in case of EX\,Lup for the lower 6\,$\mu$m flux compared to that at longer wavelengths. 
Other sources from their sample  that resemble EX\,Lup are UY\,Aur, CZ\,Tau and HP\,Tau. It is interesting to note that all four objects are binary systems with different separations.

Finally we calculated the $T_{bol}$ bolometric temperature and $L_{bol}$ bolometric luminosity of EX\,Lup following the method of Chen et al. (1995). They studied the Taurus and Ophiuchus star forming regions, and analyzed the distribution of young stars of different evolutionary stages in the $L_{bol}$ vs.  $T_{bol}$  diagram. In a later paper (Chen et al. 1997) they repeated the study for the objects of the Lupus cloud as well, including EX Lup. They found $T_{bol}=2365$\,K and $L_{bol}=0.7$\,L$_{\odot}$ based on IRAS data. Using our newly constructed SED we recalculated these two values.  Assuming $A_V=0$\,mag, our result is  $T_{bol}=1982$\,K and $L_{bol}=0.73$\,L$_{\odot}$. According to these parameters, EX\,Lup seems to be a typical classical T\,Tauri star. The location of EX\,Lup on the $L_{bol}$ vs.  $T_{bol}$ diagram implies that it is a Class II object, and its age is $3.8^{+9.2}_{-2.7}\times10^{6}$\,yr.

%\begin{comment}
\subsection{Mineralogy}
Based on the previous results that EX\,Lup is similar to classical T\,Tauri stars, in the following we assume that its circumstellar matter is distributed in a disk. 
In order to derive the dust composition in the surface layer of the disk we fitted the Spitzer IRS spectrum using the Two-Layer Temperature Distribution
(TLTD) method (Juh\'asz et al. 2009). This method assumes multi-component continuum (star, inner rim, disk midplane) 
and a distribution of temperatures to fit the source function. The silicate emission is expected to arise from the disk atmosphere.  The observed spectrum is computed as  
\begin{equation}
F_\nu = F_{\nu,\rm{cont}} + \sum_{i=1}^N\sum_{j=1}^MD_{i,j}\kappa_{i,j}
\int_{\rm{T_{atm,max}}}^{\rm{T_{atm, min}}}\frac{2\pi}{d^2}B_\nu(T)T^{\frac{2-{\rm qatm}}{{\rm qatm}}}dT, 
\label{eq:fit1}
\end{equation}
where
\begin{eqnarray}
\nonumber F_{\nu, {\rm cont}} &=&  \frac{\pi R_\star}{d^2} B_\nu(T_\star) \\
\nonumber &+& D1\int_{\rm{T_{rim,max}}}^{\rm{T_{rim, min}}}\frac{2\pi}{d^2}B_\nu(T_{\rm{rim}}){T_{\rm rim}}^{\frac{2-{\rm qrim}}{\rm qrim}}dT \\
	&+& D2\int_{\rm{T_{mid,max}}}^{\rm{T_{mid, min}}}\frac{2\pi}{d^2}B_\nu(T_{\rm{mid}}){T_{\rm mid}}^{\frac{2-{\rm qmid}}{\rm qmid}}dT.
\label{eq:fit2}
\end{eqnarray}
Here, $d$ is the distance, $B_\nu(T)$ is the Planck-function and $\kappa_{i,j}$ is the mass absorption coefficient of 
dust species $i$ and grain size $j$. The subscripts 'rim', 'atm' and 'mid' denote the quantities of the inner rim, the disk
atmosphere and the disk mid-plane, respectively. The dust species used are listed in Tab. 4.

\begin{figure}[h!]
   \centering
   \includegraphics[angle=+90, width=9cm]{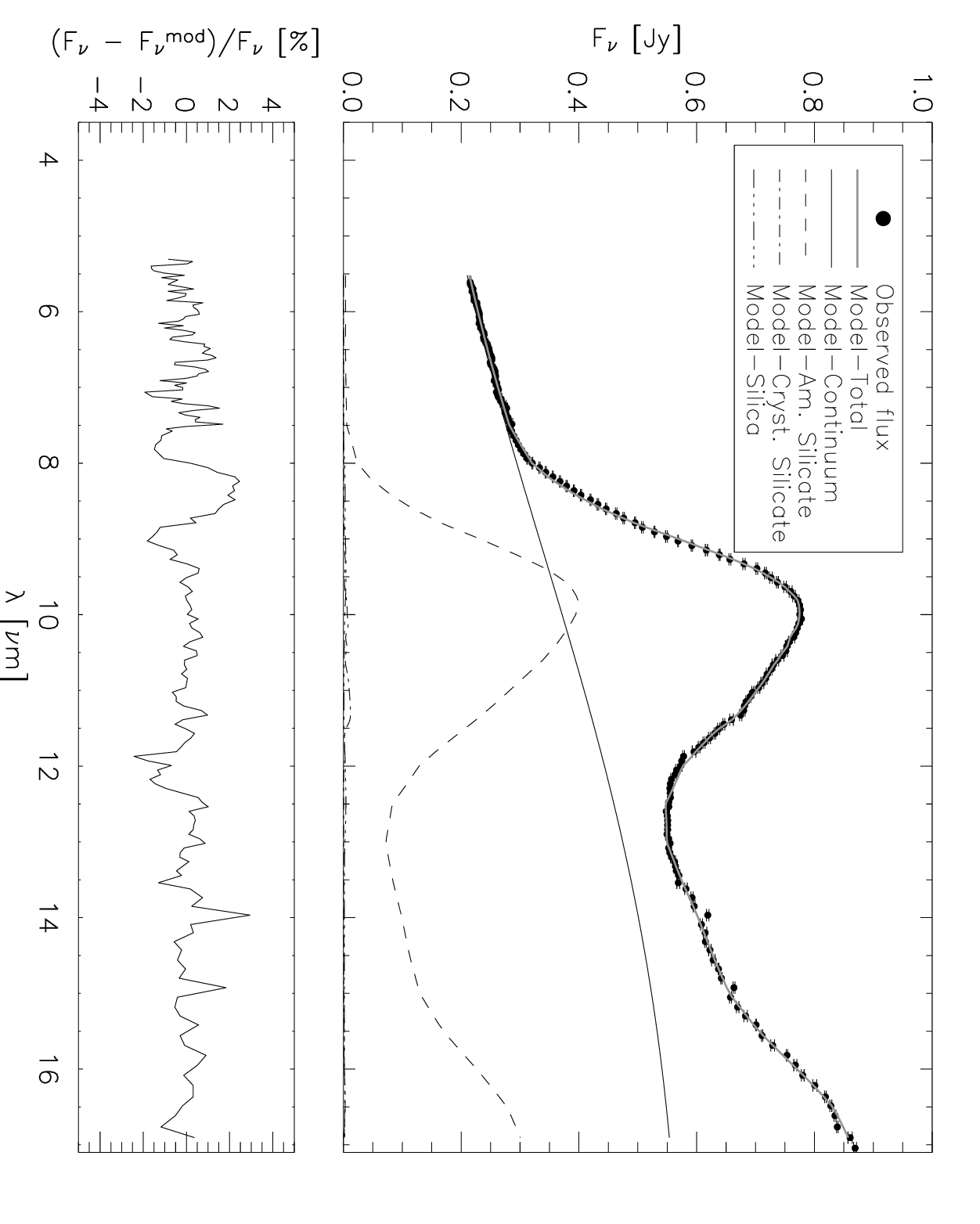}
      \caption{Fit of the observed spectrum. The spectral decomposition shows that the mid-infrared spectrum of EX\,Lup can
be reproduced by a mixture of amorphous silicates with olivine and pyroxene stoichiometry. The contribution of 
crystalline silicates derived from the spectral decomposition is below 2\%.}
         \label{fig:mineo}
   \end{figure}

In Fig.\,\ref{fig:mineo} we present the fit to the 5--17\,$\mu$m region of the Spitzer IRS spectrum. The derived dust mass fractions are given in Tab. 5. The spectral decomposition shows that the main contributors to the optically thin 10\,$\mu$m silicate emission complex are the amorphous silicates ($98.4\pm1.0\%$ in terms of mass). The mass fraction of crystalline silicates is below 2\,\%, which agrees with the value for the diffuse ISM (\cite{ref:kemper_2005}). It is interesting to note, that the mass-weighted average grain size of the amorphous silicates (0.57\,$\mu$m) in our fit is somewhat larger than typically found in the ISM. At wavelengths longer than 17\,$\mu$m the IRS spectrum of EX Lup shows only the 18\,$\mu$m feature of the amorphous silicates, no crystalline feature can be seen.

 The fitted values for $T_{\rm rim,max}$, $T_{\rm atm,max}$, $T_{\rm mid,max}$ are $1202\pm4$\,K, $1032\pm90$\,K and $846\pm4$\,K, respectively.
The fitted values for $q_{\rm rim}$, $q_{\rm atm}$, $q_{\rm mid}$ are $-2.02\pm0.01$, $-0.44\pm0.02$ and $-0.42\pm01$, respectively.
Although the $\chi^2$ of the fit is 14.5, which is far higher than unity, expected for a good fit, the average deviation from the observed
spectrum is about 1\,\%. The reason for the high $\chi^2$ can be found in the realtively high signal-to-noise ratio of the Spitzer IRS 
spectrum ($>$300 in the fitted wavelength range) and in the deficiencies of the applied dust model (optical constants and grain shape model). 
The latter is probably responsible for the differences between the 2D RT model and the observed IRS spectrum between 14 and 21\,$\mu$m.

Given the fact that EX\,Lup is a young eruptive star, one would expect to observe an increased value of crystallinity in the mid-infrared features, compared to 'normal' T\,Tauri stars. Since the only requirement of the crystallization is the high temperature, the enhanced irradiation luminosity and viscous heating during the outbursts should lead to rapid crystallization in the disk. \cite{ref:quanz_2007} (2007), in accordance with our results, reported the lack of crystalline emission and the presence of emission from larger grains ($a>0.1$\,$\mu$m)  in the mid-infrared spectra of FU Orionis objects. 
They hypothesized that the reason for the lack of crystals can be twofold. One can be the replenishment of  the dust content 
of the disk atmosphere (where the mid-infrared features originate) by pristine dust from an infalling envelope of the FU Ori object.  Another possible explanation 
 is strong vertical mixing in the disk which transports the crystals into deeper layers of the disk where they cannot be detected
any more by mid-infrared spectroscopy. Out of these two speculative scenarios the latter is favorable in case of EX\,Lup since,
to our knowledge, there is no infalling envelope around the source.

\begin{table}[!h]
\caption{Overview of dust species used in fitting the 5$-$17\,$\mu$m spectrum. For each component we specify
 its lattice structure, chemical composition, shape and reference to the 
 laboratory measurements of the optical constants. For the homogeneous
 spheres we used Mie theory to calculate the mass absorption/scattering coefficients. For the inhomogeneous
 spheres, we used the distribution of hollow spheres (\cite{ref:min_2005}), to simulate
 grain shape deviating from perfect symmetry. References: (1) \cite{ref:dorschner_1995},
 (2) \cite{ref:servoin_1973}, (3) \cite{ref:jaeger_1998}, (4) \cite{ref:henning_1997}.}
\begin{center}
\begin{tabular} {l@{\hspace{2.5 mm}}l@{\hspace{2.5 mm}}l@{\hspace{2.5 mm}}l@{\hspace{2.5 mm}}l}
\hline\hline\\
 Species                & State & Chemical            & Shape & Ref \\ 
                        &       &  formula             &      &   \\  \hline
 Amorphous silicate     & A     &  MgFeSiO$_{4}$       & Homogeneous           & (1)  \\
 (olivine stoichiometry)&       &                     &  sphere             &  \\
 Amorphous silicate     & A     &  MgFeSi$_{2}$O$_{6}$ & Homogeneous     & (1)  \\
 (pyroxene stoichiometry)&      &                     &  sphere            &  \\
 Forsterite             & C     &  Mg$_{2}$SiO$_{4}$   & Hollow sphere & (2)   \\
 Clino Enstatite        & C     &  MgSiO$_{3}$         & Hollow sphere & (3) \\
 Silica                 & A     &  SiO$_{2}$           & Hollow sphere & (4)  \\
\hline
\end{tabular}
\end{center}
\label{tab:dust_species}
\end{table}

\begin{table}[!h]
\caption{Fitted dust composition to the Spitzer IRS spectrum. We only show mass fractions that are larger
than 0.1\,\%.}
\begin{center}
\begin{tabular}{lccc}
\hline\hline
Dust species & \multicolumn{3}{c}{Mass fraction [\%]}\\
                        & 0.1\,$\mu$m      & 1.5\,$\mu$m     & 6.0\,$\mu$m \\
\hline\\
Am. Silicate       	& 65.6 $\pm$ 0.5   &    --           &     --       \\
(Olivine type)&&&\\
Am. Silicate 	        & --               & 32.80 $\pm$ 0.5 &     --        \\
(Pyroxene type)&&&\\
Forsterite              &   0.7 $\pm$ 0.08 &  --             &     --        \\
Enstatite 	        &   --             & 0.5 $\pm$ 0.3   &  0.4 $\pm$ 0.4 \\
Silica		        &   --             &     --          &  --            \\
\hline	
\end{tabular}
\end{center}
\label{tab:fit_dustcomp}
\end{table}

%\end{comment}

\subsection{Modeling}

In this section we perform a detailed modeling of the EX Lup system in order to derive the geometry of its circumstellar environment. We fit the data points presented in Fig.\,1, except the mid-infrared domain where, due to the intrinsic variability and the different quality of the data, we considered the Spitzer measurements only: the IRAC and IRS observations form a quasi-simultaneous high quality data set covering the $3.6-38$~$\mu$m wavelength range. Moreover, these measurements are close in time to the recent 2008 eruption and thus document the pre-outburst conditions of the system.

For the stellar parameters (radius $R_*$, effective temperature $T_*$ and mass $M_*$) we adopted the values used by \cite{ref:grasvelazquez_2005} (2005). These parameters were  fixed during the modeling (Tab. 6 lists all input parameters marking differently the fixed and the optimized ones).  For the distance of the star we adopted 155\,pc, the mean distance of the Lupus Complex as measured by the Hipparcos (\cite{ref:lombardi_2008}). Note that the unknown location of EX\,Lup within the complex introduces an additional uncertainty of about 30 pc. For the value of extinction we assumed $A_V=0$ mag (Sect.\,3.2). 

\begin{figure}[h]
   \centering
   \includegraphics[ width=7cm]{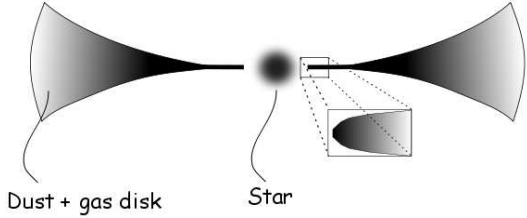}
      \caption{Schematic picture of the geometrical structure of the model. The figure is not to scale.}
         \label{moricka}
\end{figure}

We used the Monte  Carlo radiative transfer code \emph{RADMC} (\cite{ref:dullemond_2004}) combined with \emph{RAYTRACE}\footnote{for details see: http://www.mpia.de/homes/dullemon/radtrans/radmc/}. The circumstellar environment is supposed to be axially symmetric, so we use a two-dimensional geometry in polar coordinates ($r,\theta$). As mentioned in Sects.\, 2.1 and 4.1 we had no indication for the presence of an envelope around EX\,Lupi, thus our system consists only of a central star and a dusty disk. For the structure of the disk we assume the density profile as follows:
\[
\rho_{disk}(r,z) = \frac{\Sigma (r)}{h(r)\sqrt{2\pi}} \exp{\left\{-\frac{1}{2}\left[ \frac{z}{h(r)}\right]^2\right\}},
\]
where $r$ and $z$ are the radial and vertical coordinates respectively, $\Sigma(r)=\Sigma_{disk}\left({r\over r_{disk}}\right)^{-p}$ is the surface density. The 'disk' subscript denotes values of the given parameter at the outer radius. The scale height $h(r)$ increases with radius as:
\[
{h(r)\over r}=\frac{h_{disk}}{r_{disk}}\left(\frac{r}{r_{disk}}\right)^{\alpha_{fl}},  
\]
 where $\alpha_{fl}$ is the flaring index. A schematic picture of the used geometry is presented in Fig.~\ref{moricka}. 

The temperature distribution is determined by the heating sources: the central star, for which we used a Kurucz model atmosphere (\cite{ref:castelli_2003}) and the heated dust grains emitting blackbody radiation. We used a passive disk and did not consider accretion, due to the low accretion rate in quiescence (Sect.\,3.3). The accretion luminosity is $<$ 1\% of the stellar luminosity, thus its contribution is negligible compared to direct irradiation coming from the central star.  After the calculation of the temperature distribution, the SED  of the system is produced at an inclination angle $\vartheta$ with the ray-tracer.

We used the dust composition derived from the fitting of the Spitzer IRS spectrum (Sect.\,4.2), but excluded dust species whose contiburtion to the total mass is  $\le$ 1\%. This way, the dust model used in \emph{RADMC} contained only amorphous silicates of olivine and of pyroxene types with a mass ratio of 2:1. The size of the dust grains were 0.1\,$\mu$m and 1.5\,$\mu$m for olivine and pyroxene stoichiometry, respectively. Besides these, we added 20\,\% amorphous carbon with a grain size of 0.1\,$\mu$m. The mass absorption coefficients of amorphous carbon were calculated using Mie-theory from the optical constants of \cite{ref:preibisch_1993} (1993).

\begin{table}[h!]
\caption{Parameters used in the best-fit model. Those in italics were adopted from the literature and kept fixed during the modeling.}
\centering
\begin{tabular}{l r@{\hspace{.7 mm}} l}
\hline\hline
Parameters&\multicolumn{2}{c}{Fitted value}\\\hline
\textbf{System Parameters}                  &         &\\
\hspace{0.5 cm}\emph{Distance} ($d$)                               & 155    &pc\\
\hspace{0.5 cm}Inclination ($\vartheta$)                    &20      &$^{\circ}$\\
\textbf{Stellar Parameters}                  &         &\\
\hspace{0.5 cm}\emph{Temperature} ($T_{star}$)                     &3800    &K\\
\hspace{0.5 cm}\emph{Mass}  ($M_{star}$)                           &0.6     & $\rm{M}_{\odot}$ \\
\hspace{0.5 cm}\emph{Radius} ($R_{star}$)                           &  1.6     &$\rm{R}_{\odot}$\\
\hspace{0.5 cm}\emph{Visual extinction} ($A_V$)                    &  0      &mag\\
\textbf{Circumstellar Disk Parameters}       &        &\\
\hspace{0.5 cm}Inner radius of dusty disk ($r_{in,disk}$)    & 0.2      & AU\\
\hspace{0.5 cm}Outer radius of dusty disk ($r_{disk}$)   &150  &AU\\
\hspace{0.5 cm}Scale height ($\frac{h_{disk}}{r_{disk}}$)               & 0.12     &\\
\hspace{0.5 cm}Flaring index ($\alpha_{fl}$)                      & 0.09     &\\
\hspace{0.5 cm}Exponent of radial density profile ($p$)&-1.0\\
\hspace{0.5 cm}Total mass ($M$)                             &0.025     &$\rm{M}_{\odot}$\\
\hline
\end{tabular}
\label{tab:para}
\end{table}

\begin{figure}[t]
   \centering
   \includegraphics[angle=0, width=9cm]{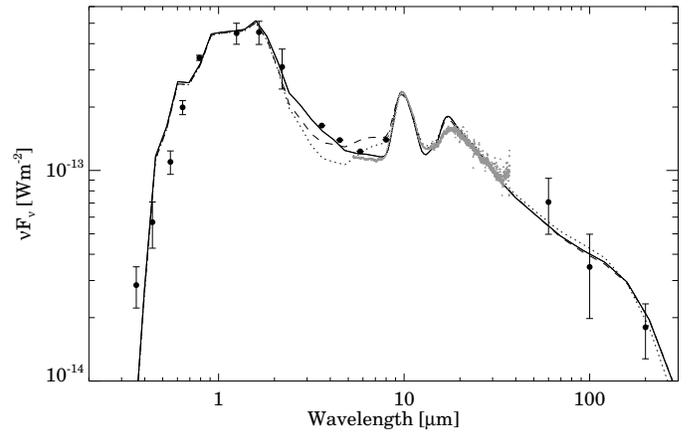}
      \caption{Spectral energy distribution of EX\,Lup. The solid line shows our best fit model with the rounded inner wall, while the dashed and dotted lines correspond to the best models using a sharp inner edge. The plotted error bars mark the range of observed fluxes at different epochs, taking into account also the individual measurement uncertainties (filled dots correspond to the middle of the range). Error bars on the IRAC points are smaller then the size of the symbols.}
         \label{fig:modelsed}
\end{figure}

We could fit reasonably most of the measured data points by adopting the geometry described above. In the best model of this type we had to move the inner radius of the dusty disk out to 0.5\,AU, significantly exceeding the dust sublimation radius (at T=1500\,K) of less than 0.1 AU. However, in the 3$-$8\,$\mu$m wavelength range this model underestimated the measured points (Fig.\,\ref{fig:modelsed}, dotted line). Using smaller values for the inner radius, we could improve the fit of the 3$-$5\,$\mu$m range, but then it was not possible to reproduce the measured fluxes at longer wavelengths (Fig.\,\ref{fig:modelsed}, dashed line). Reducing the inner radius below 0.2 AU  the model fluxes even in the near-infrared wavelength range became too high. In order to solve this problem, we introduced a rounded inner rim to the disk (see inset in Fig.\,\ref{moricka}), thus instead of having a sharp inner edge with a high wall we decreased the disk height to $\frac{h_{disk}}{r_{disk}}= 0.05$ at the beginningof the disk and at the same time moved the inner radius inward. Behind 0.6\,AU we left the structure of the disk unchanged. With this modification we could obtain a fit that is in good agreement with all infrared observations. Nevertheless, the inner radius of 0.2\,AU used in the best model is still beyond the sublimation radius.

Our best-fit model is presented in Fig.~\ref{fig:modelsed} with a solid line, the corresponding parameters are listed in Tab.~\ref{tab:para}. 
In the literature we could not find any constraints for the inclination of the system and its value is not well defined by the model either. The disk is definitely seen closer to face-on then edge-on, and we obtained our best fit using a value of $20^{\circ}$, though fits for inclinations between $0^{\circ}$ (face on) and $~40^{\circ}$ are very similar. The outer radius of the circumstellar disk  is typical for young systems, while the disk is massive among T Tauri disks. For the value of the flaring index $\alpha_{fl}=0.09$ gave the best result, which means that the disk flares only very modestly.  The scale height is $h=12$ AU at the outer radius of the disk, which is slightly higher than the value typical for T Tauri disks. 
The exponent of the radial surface density profile is $p=-1.0$.

\section{Summary and conclusions}

We characterized the quiescent disk of EX\,Lupi and investigated whether it is the structure of the circumstellar environment which makes 
it an atypical, eruptive young stellar object. Our main findings are the following:
\begin{itemize}
\item During quiescent phase there is indication for an intrinsic variability of less than 25\% in the optical--mid-infrared wavelength regime.
\item Our new spectra is consistent with the classification of EX\,Lup as an M type star. Based on the H$\alpha$ and Pa$\beta$ spectral lines, we derived a very low quiescent accretion rate of $\sim 4\times 10^{-10}$\,M$_{\odot}$/yr.
\item  In general the shape of the SED is similar to those of typical T\,Tau stars, but above 7\,$\mu$m EX\,Lup is brighter than the Taurus median by a factor of $\sim$2.5. The relative flux contribution from shorter and longer wavelengths is a parameter which may distinguish EX\,Lup from the majority of classical T\,Tau stars.
\item  The 10\,$\mu$m silicate feature of EX Lup can be well reproduced by
  amorphous silicates with olivine and pyroxene stoichiometry, no
  crystalline silicates were found.
\item A modestly flaring disk model with a total mass of 0.025 M$_{\odot}$ and with inner and outer radii of 0.2 and 150 AU, respectively, is able to reproduce the observed SED. The radius of the inner hole is larger than the dust sublimation radius.
\end{itemize}

The existence of this dust free inner hole points to a clearing mechanism, for which several explanations could be invoked:
(1) \cite{ref:eisner_2007} claimed that low mass young stars with small accretion luminosities tend to have inner disk radii larger than the sublimation radius,  probably due to effects of the magnetic field. Using the relationship in \cite{ref:eisner_2007} and taking 0.2\,AU as the magnetospheric inner radius and an accretion rate of $4\times 10^{-10}$\,M$_{\odot}$/yr EX\,Lupi should have a magnetic field strength of 2.3\,kG, which is typical for T Tauri stars (\cite{ref:johnskrull_2007}).\\ 
(2) Binarity might also be responsible for clearing up regions of the disk. In case of EX\,Lup binarity was studied by several authors. \cite{ref:ghez_1997} used high angular resolution techniques in order to find wide components, and detected no companion of EX\,Lup between 150$-$1800\,AU separation. \cite{ref:bailey_1998} detected no companion between $1-10$\,AU. \cite{ref:melo_2003} claimed that EX\,Lup is not a spectroscopic binary,  and similarly Herbig (2007) concluded the same based on spectroscopic data measured by the Keck telescope. \cite{ref:guenther_2007} performed an eight year long radial velocity monitoring program, but they could not detect binarity in case of EX\,Lupi either though they only had 3 spectra. Nevertheless, we note that  if the inclination of EX\,Lupi is indeed close to a face-on geometry it makes detection of binarity difficult.\\
(3) Inner gaps are characteristic of disks in transitional phase between the Class\,II and Class\,III stages. These transitional disks, however -- unlike EX\,Lup -- exhibit no excess over the photosphere in the near-infrared wavelength range. Furlan et al. (2006) suggested that IRAS\,04385+2550, an object whose SED is similar to that of EX\,Lup, could be in a state preceding the transition. Adopting this explanation, EX\,Lup might also be in a pre-transitional state. 
\\
(4) Alternatively with such a low accretion rate photoevaporation by EUV radiation may also contribute to the clearing of this innermost region (\cite{ref:gorti_2009}).
\\
\textbf(5) It cannot be excluded that the large inner hole is a pheanomenon in connection with the eruptive behaviour, although details of this connection are unclear yet.

Our detailed modeling of the quiescent disk structure of EX\,Lup (Sect.\,4.2) may be a good basis for studying the physical changes related to the 2008 eruption. Assuming that the outburst is due to temporarily increased accretion, one could scale up the accretion rate in our quiescent model, and check whether this new model would reproduce the outburst SED. If this strategy fails to provide a sufficiently good fit to the outburst data, it may be a hint for geometrical restructuring of the circumstellar environment during the outburst. 

In the present work one of our main goals was to identify atypical features in the circumstellar structure of EX\,Lup which may explain its eruptive nature. The inner disk hole revealed by our modeling is an  unexpected result in this sense and comparison of EX \,Lup with other EXors at infrared wavelengths would  be important. It may answer the question whether an inner gap in the dusty disk is characteristic of the EXor phenomenon thus the hole is connected to the eruption mechanism, and we could learn to which extent is EX\,Lup a good representative of eruptive stars.

\begin{acknowledgements}
The work was supported by the grants OTKA T\,49082 and OTKA K\,62304 of the Hungarian Scientific Research Fund. The authors are grateful for C. P. Dullemond for kindly providing \emph{RAYTRACE} and also for Jeroen Bouwman for providing his routines for the Spitzer IRS data reduction. The research of \'A. K. is supported by the Nederlands Organization for Scientific Research. They also thank for observational data from the AAVSO International Database contributed by observers worldwide.
\end{acknowledgements}

\end{document}